\documentclass{article}
\begin{document}
\title{Energy Loss by Gravitational Viscosity}
\author{Ernst Fischer\footnote{Auf der H\"ohe 82, D-52223 Stolberg, Germany}}
\date{e.fischer.stolberg@t-online.de}
\maketitle

\begin{abstract}
Due to Lorentz invariance of General Relativity gravitational interaction is
limited to the speed of light. Thus for particles, moving within a matter
field, retardation leads to loss of energy by emission of gravitational
radiation. This 'gravitomagnetic' effect, applied to motion in homogeneous
mass filled space, acts like a viscous force, slowing down every motion in
the universe on the Hubble time scale. The energy loss rate exactly equals
the red shift of photons in an expanding universe, thus showing the
equivalence of wavelength stretching in the wave picture and energy loss in
the photon picture. The loss mechanism is not restricted to an expanding
universe, however, but would also be present in a static Einstein universe.
\end{abstract}


\section{Introduction}
Today the theory of general relativity (GRT) is accepted as the correct
description of gravitation, but due to its non-linear character and the
complicated mathematical formalism practical applications have been rather
limited and, wherever it appears acceptable, the much simpler formalism of
Newtonian theory is used, which appears as a good approximation in the limit
of weak fields and low velocities. One of the most important differences from
Newtonian gravity is the Lorentz invariance of GRT, following from the
requirement that gravitational interaction should be independent from any
preferred reference system other than the complete universe.

One consequence of this requirement is the fact that gravitational
interaction is not instantaneous but limited to the speed of light, leading
to similar radiation effects as we know them from electromagnetic
interaction. Changes of the matter or energy distribution lead to changes of
the metric, expanding into space by the speed of light. Though this effect
has been already discussed by Einstein in his famous quadrupole formula, it
remained of little practical interest up to the last two decades. The first
observation, which could be attributed to this gravitational radiation, was
the detection that the frequency of double pulsars was slowing down exactly
according to GRT (Weisberg and Tailor (1984) \cite{weisberg}). Besides that,
several attempts have been made to detect gravitational waves directly, which
should be emitted by cosmic catastrophes such as collisions of black holes or
other huge matter concentrations. An overview of the theoretical models
developed to describe such events has been given by Poisson (2004)
\cite{poisson}.

But while these activities concentrate on processes, which produce very
strong deformations of space-time, there should be also effects in the weak
field limit, which cannot be detected directly, but which may be important,
when they accumulate over cosmic distances or times. They may influence the
development of cosmic structures like the formation of galaxies or clusters.

The primary effect, which makes up the difference between Newtonian theory
and GRT is the finite velocity of gravitational interaction. Thus we will try
to investigate such processes in the limit of Newtonian physics, but with the
modification that the interaction velocity is given by the speed of light,
just as we are accustomed from the theory of electromagnetic interaction. In
analogy to the magnetic effects observed in the interaction of moving
electrical charges, these deviations from Newtonian theory are mostly called
'gravitomagnetic' and their existence has been proved with high accuracy in
lunar and planetary ranging observations (see e.g. Nordtvedt (2003)
\cite{nordtvedt}).

In electromagnetic theory of moving charges static potentials have to be
replaced by the retarded Li\'{e}nard-Wiechert potentials. That this
retardation has an effect on the motion of particles also in gravitational
theory can be immediately recognised from a simple thought experiment. We
consider a particle moving on a line between two equal masses. When the
particle at time $t$ is just at equal distance $r$ from both masses, the
gravitational force at this moment is determined by the distance at the time
$t-r/c$. That means that the distance to the mass in direction of the motion
is increased and the distance to the other mass is reduced. Thus the particle
feels a force, which is directed opposite to the direction of motion, thus
reducing its momentum and energy of motion.

Of course, if we try to sum up the forces exerted on a test particle by the
total matter in a Euclidean universe, we are confronted with the infinity
problems, well known from the Olbers paradox. But also here general
relativity supplies us with a remedy, or better to say, with two possible
solutions. One is the assumption, favoured by main stream physics of today,
that the universe is expanding and thus, even if the size of the universe is
infinite, interaction is limited to that fraction of matter, which can be
causally connected to the test particle. But possible is also the other
explanation, originally proposed by Einstein, the assumption that space is
curved and thus of finite size and matter content. In the sequel we will
consider both possibilities and show that they lead to similar results.

\section{Energy loss in expanding space}
To demonstrate, how the finite speed of gravitational interaction affects the
energy balance of moving particles, as a toy model we consider the motion of
a test mass in a universe, expanding at a constant rate, but keeping the
density constant by some creation process similar to that proposed by Hoyle
et al. (1993) \cite{hoyle}. This does not mean that this model appears more
attractive than the presently favoured 'concordance model'. But it allows a
mathematically very simple description of the effects, as the relevant
quantities, matter density and expansion velocity, do not depend on time.

Starting with Newtonian physics, the gravitational potential at some point
generated by masses $m_i$ at distances $r_i$ is given by
\begin{equation}
U=\sum_i \frac{Gm_i}{r_i}
\end{equation}
($G$ is the gravitational constant). But if gravitational interaction is
limited by the speed of light, similar to electromagnetic interaction, if the
position of the masses changes with time, we have to use retarded potentials
analog to the Li\'{e}nard-Wiechert potentials. Instead of the distance at the
local time $t$, we have to insert the distance at time $t-\tau$, where $\tau$
is the running time of the signal $\tau=r_i/c$.
\begin{equation}
U^*(t)=\sum_i \frac{Gm_i}{r_i(t-\tau)}.
\end{equation}
In the static case this does not change the result, but for a moving particle
the retardation parameter $t-\tau$ changes with time. As a result there
occurs an additional gradient of the potential, leading to a retarding force
on the particle.

We should stress here that analogous to electromagnetic interaction
retardation does not lead to aberration effects, as Lorentz invariance
requires that there exists no preferred reference frame. Thus interaction can
depend only on the scalar retarded potential, not on the direction of forces
as determined from Euclidean geometry. A detailed discussion of this fact has
been given by Carlip (2000) \cite{carlip}. While a test particle, linearly
moving in the field of homogeneously distributed masses, feels a retarding
force, as the distance to the individual masses changes with time, in
circular motion of an isolated two-body system there is no retarding force.
As the distance of the masses does not change, angular momentum is conserved.
Thus in the solar system deviations from Newtonian physics occur only as tiny
corrections of higher order in $v/c$ due to rotation, tidal effects and
eccentricity of the moving bodies. In linear motion there exists an effect of
first order in $v/c$, however.

Let us assume a particle moving at velocity $v$ in positive x-direction of
Euclidean space. Using a comoving coordinate system, we can regard all the
other masses as moving with respect to it with speed $-v$. Thus their
distance changes with time by
\begin{equation}
\frac{dr_i}{dt}= \frac{dr_i}{dx}\cdot\frac{dx}{dt}=\frac{x_i}{r_i}\cdot\frac{dx}{dt}
=-v \frac{x_i}{r_i}
\end{equation}
In the comoving system there is a gradient of the retarded potential, which
is felt by the particle as a retarding force. It can be expressed by a power
series
\begin{equation}
\frac{dU^*}{dx}=\sum_i Gm_i\frac{d}{dx}\sum_{n=0}^\infty \frac{d^n}{d\tau^n}\left(\frac{1}{r_i}\right)
\frac{\tau ^n}{n!}.
\end{equation}
Restricting to the linear term of the series expansion with $\tau=r_i/c$ we
get
\begin{eqnarray}
\frac{dU^*}{dx}&=&\sum_i Gm_i\frac{d}{dx}\left(\frac{1}{r_i}+\frac{vx_i}{r_i^3}\frac{r_i}{c}\right)
\nonumber\\ &=& \sum_i
Gm_i\left[-\frac{x_i}{r_i^3}+\frac{v}{c}\left(\frac{1}{r_i^2}-\frac{2x_i^2}{r_i^4}\right)\right]
\end{eqnarray}
We can now apply this formula to the motion of particles in an homogeneous
matter filled space of density $\varrho$. In this case the mass in a toroidal
volume element at distance $r_i$ is $m_i=\varrho dV_i=2\pi\varrho\,
r_i^2\cos\vartheta
\,d\vartheta \,dr$, where the projected distance in the direction of motion is
expressed by $x=r\sin\vartheta$. Then integration of the potential gradient
induced by the matter is
\begin{equation}
\label{int}
\frac{dU^*}{dx}=2\pi G \varrho
\int_0^{r_{max}}\int_{-\pi/2}^{\pi/2}Y(r,\vartheta)\,r^2\cos\vartheta\, d\vartheta\, dr
\end{equation}

\begin{equation}
\textrm{with}\quad Y(r,\vartheta)=\left[
\frac{sin\vartheta}{r^2}+\frac{v}{c}\left(\frac{1}{r^2}-\frac{2\sin^2\vartheta}{r^2}\right)
\right]
\end{equation}

The quantity $r_{max}$ has been introduced to express that integration has to
be limited to the causal sphere, that means, to the volume range, which can
interact, when this interaction is limited to the speed of light. Thus, if
space is expanding at a rate given by the Hubble constant $H$, the limit is
$r_{max}=c/H$.

From symmetry considerations it is immediately clear that the first term of
the integral is zero, but the second term gives a non-vanishing contribution
proportional to $v/c$:
\begin{equation}
\frac{dU^*}{dx}=2\pi G \varrho\, r_{max}\cdot\frac{v}{c}\left(\sin\vartheta -\frac{2}{3}
\sin^3\vartheta\right)_{-\pi/2}^{\pi/2}
\end{equation}
corresponding to a force on a particle of mass $m$
\begin{equation}
F=-m\frac{dU^*}{dx}=-\frac{4\pi}{3} G \varrho\, r_{max}\cdot\frac{mv}{c}
\end{equation}
Introducing $r_{max}=c/H$ there is a loss of momentum $p$ proportional to the
actual momentum
\begin{equation}
\frac{dp}{dt}=-\frac{4\pi}{3} \frac{G \varrho}{H} \cdot p,
\end{equation}
and correspondingly the loss of kinetic energy is
\begin{equation}
\frac{dE}{dt}=\frac{d}{dt}\left(\frac{p^2}{2m}\right)=-\frac{8\pi}{3}
 \frac{G \varrho}{H} \cdot E.
\end{equation}
As in an expanding universe the case of zero curvature corresponds to the
critical density $\varrho_{cr}=3H^2/(8\pi G)$, this energy loss corresponds
exactly to that of photons $dE/dt=-HE$. That means that every moving particle
in a mass filled universe of critical density experiences the same energy
loss as we ascribe it to photons and explain it by stretching of wavelength
by the expansion of space. Or looking at it the other way round: What we
explain by stretching of wavelengths in a pure wavelike picture of
electromagnetic radiation, may as well be considered as part of a general
energy loss mechanism of all matter, if we regard photons as individual
entities, which have inertia $h\nu/c^2$ and feel gravity in the same way as
massive particles.

The situation is much the same as in the experiments by Pound and Rebka
(1960) \cite{pound} to measure gravitational red shift. The results can be
explained in a geometrical way by time dilatation, caused by spacetime
deformation by the surrounding matter, but as well by the difference of the
gravitational potential between the points of emission and detection of
radiation. Equivalence of both pictures results from conservation of energy,
which is valid, when potential energy is included.

Of course, that extending the quasi-Newtonian approximation to photons is
allowed, has not been proved. But the fact that the numerical values are
exactly equal and that energy conservation should be valid, independent of
the nature of the moving mass or quantum, is an indication that the energy
loss effect is a common phenomenon to massive and massless particles.
Otherwise the principle of equivalence, which is a corner stone of GRT, would
be violated. Mass filled space appears as a kind of viscous medium, in which
energy of all moving particles is dissipated to the gravitational potential.

\section{Energy loss in curved space}
In the last section we have shown that the cosmic energy loss mechanism must
not necessarily be attributed to an expansion of space. The basic proposition
is the finite speed of gravitational interaction, which is a consequence of
the Lorentz invariance of the basic GRT equations. It is this Lorentz
invariance, which requires that any homogeneous space solution of the basic
equations must be either expanding or spatially curved. Energy loss by
gravitational viscosity, as we have called it, must be present, too, if space
is not expanding. We only have to find another explanation, why gravitational
interaction is limited to a finite amount of matter, if this amount is not
limited by causal connection.

There is of course such a possibility, when the total amount of matter in the
universe is limited, as it was proposed in Einstein's first introduction of
GRT. He proposed that the universe is static, but has a positive curvature,
so that the total matter content is limited. We can easily adjust the
derivation of cosmic energy loss to this case, we only have to change the
size of the volume elements and the limits of integration.

We consider a homogeneous spherical universe with a positive radius of
curvature $R$. As in Euclidean space we assume that gravity acts along the
geodesic lines and that the strength of interaction decreases with the square
of the length measured along these lines. Denoting this distance by $r=R
\cdot\varphi$, the main difference compared to Euclidean space is the reduction
of the volume element at distance $r$ by the factor
$(\sin\varphi/\varphi)^2$. There exists, of course, no limitation to the
length of the geodesic lines, so that the value of $\varphi$ may extend to
infinity. But due to the limited size of the volume elements the value of the
integral remains finite. Thus in eq.(\ref{int}) we only have to replace the
integral
\begin{equation}
\int_0^{r_{max}}dr= r_{max}\quad\textrm{by}\quad R \int_0^{\infty}\frac{sin^2\varphi}{\varphi^2}
\, d\varphi=\frac{R \pi}{2},
\end{equation}
leading to an energy loss
\begin{equation}
\label{estat}
\frac{dE}{dt}=-\frac{4\pi^2}{3} \frac{G \varrho R}{c} \cdot E.
\end{equation}
The relation between the radius of curvature and the density of matter in the
static Einstein universe is given by $R=\sqrt{c^2/(4\pi G \varrho)}$. Thus,
if we again identify the energy reduction factor in eq.(\ref{estat}) with the
Hubble constant $H$, the density of matter is related to $H$ by
\begin{equation}
\varrho=\frac{9H^2}{4\pi^3G},
\end{equation}
which differs not much from the critical density in an expanding universe.
The factor is $\varrho /\varrho_{cr} = 6/\pi^2$. From the viewpoint of red
shift or global energy loss there is no reason, to prefer one model against
the other.
\section{Discussion}
As has been shown in the last sections, the gravitomagnetic effects inherent
to GRT or, as we call it in the context of global effects, 'gravitational
viscosity' can explain observed red shift with and without any expansion of
space as well. To prove, if the general energy loss mechanism on cosmic scale
really exists, our possibilities are rather limited. To prove the existence,
we would have to follow the path of moving particles over very long periods
of time and to know the initial energy of motion exactly. But unfortunately
the only particles, for which the initial energy is known from the emission
process, are photons - at least if we assume that the laws of quantum
electrodynamics have hot changed with time.

Observations within the solar system have unambiguously shown, that the
gravitomagnetic effects exist. The planetary and lunar ranging experiments
(see \cite{nordtvedt}), by which the distance from earth to moon could be
measured down to an accuracy limit of cm, have demonstrated the validity of
GRT in an impressive way. The effect of global energy loss is just at the
accuracy limit of these measurements. It would change the distance to the
moon by about 3 cm per year. But there are so many perturbating effects of
the same or higher magnitude within the solar system that the observed
changes cannot be unambiguously attributed to the global loss mechanism. And
even if we succeed to isolate the global change from all the other
perturbations, it remains the question, if it is due to expansion or
curvature.

What we need to decide this question, are measurements, which do not relay on
red shift or similar energy loss, but only on geometric observations like
angular size or mean number density of distant galaxies. But also the
interpretation of these measurements is hampered by the possibility that the
properties of galaxies may have changed with time. Thus a conclusive decision
is still missing. But an increasing number of observations shows that the
most distant galaxies look quite similar as those nearby, casting doubts on
the assumption that a strong development in size or properties has taken
place since the light has been emitted, which is reaching us now.

There is one indirect hint, however, that gravitational viscosity, at least
by part, is responsible for the observed energy loss of moving matter and
consequently also for red shift of photons. Without assuming some dissipation
of kinetic energy, which matter acquires while contracting into galaxies or
clusters, formation of these structures cannot be explained. Gravitational
viscosity supplies this dissipation effect. The fact that we observe galaxies
as spirals and not as circular disks, can easily be attributed to the loss of
angular momentum during formation of structures on the Hubble time scale.

Thus, if we take GRT serious as the correct theory of gravitation, we should
do it with all its consequences. We should not regard Doppler effect in an
expanding universe as the only possible explanation of observed red shifts.
Taking into account energy loss by gravitational viscosity, may help us to a
better understanding of the formation phenomena of large scale cosmic
structures and it may even question present days assumptions on the age of
the universe, derived from observed red shift.

\label{lastpage}
\end{document}